\documentclass[prb,twocolumn]{revtex4}
\usepackage{pstricks}
\usepackage{amssymb,amsmath,array,graphicx,epsfig,color}
\begin{document}
\newcommand{\td}{\textup{d}}  
\def\Pcm#1{{\mathcal{#1}}}
\newcommand{\del}{\partial}
\def\eqref#1{(\ref{#1})}
\def\er#1{eqn.\eqref{#1}}
\def\nn{\nonumber}

\title{ Three results on weak measurements.} 
\author{ N.D. Hari Dass }
\email{dass@tifrh.res.in }
\affiliation{TIFR-TCIS, Hyderabad 500075 }

\maketitle

\hrule\vspace{2mm}
{\noindent\bf
Three recent results on weak
measurements are presented. They are: i) repeated measurements on a single copy can not provide any information 
on it and further, that in the limit of very large such measurements, weak measurements
have exactly the same characterstics as strong measurements, ii) the apparent non-invasiveness of weak measurements is 
\emph{illusory} and they are no more advantageous than
strong measurements even in the specific context of establishing Leggett-Garg inequalities,
when errors are properly taken into account, and, finally, iii) weak value measurements are optimal, in the precise sense of Wootters and Fields, when the
post-selected states are mutually unbiased with respect to the eigenstates of
the observable whose weak values are being measured. Notion of weak value coordinates for state spaces
are introduced and elaborated.
}
\vspace{2mm}\hrule\vspace{2mm}

\noindent
{\bf Keywords:} Projective, Weak , Repeated weak , Non-invasive, Optimal.  

\section{Quantum Measurements}
\begin{figure}[htp!]
  \centering
  \includegraphics[width=1.5in]{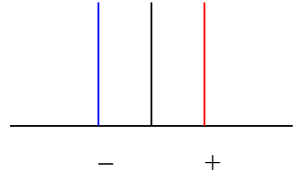}
  \caption{Measurement interaction of eqn.(\ref{eq:measint}).}
\label{fig:measint0}
\end{figure}

The chief ingredients for a quantum measurement on a quantum system are i) an appropriate apparatus, with well defined 
pointer
states $P_i$ (these, in the present folklore, are to be determined by suitable apparatus decoherence processes), and an 
appropriate
measurement interaction ${\cal M}$ between the system and the apparatus. The latter is determined by the observable of the 
system to be measured. A point to be emphasised is that the \emph{same} measurement interaction can be used both for 
the strong,
projective measurements, as well as for the so called \emph{weak measurements}. For example, for qubit measurements, this 
can be taken to be (A,S are for apparatus and system,respectively, and $P_i$ are the pointer-states of the apparatus): 
\begin{eqnarray}
\label{eq:measint}
& &|P_i\rangle_A\otimes|\uparrow\rangle_S\xrightarrow{{\cal M}}\,|P_{i+1}\rangle_A\otimes|\uparrow\rangle_S\nonumber\\
& &|P_i\rangle_A\otimes|\downarrow\rangle_S\xrightarrow{{\cal M}}\,|P_{i-1}\rangle_A\otimes|\downarrow\rangle_S
\end{eqnarray}
This is sybolically depicted in Figure.(\ref{fig:measint0}) where the central line denotes the pointer state $P_i$, 
and those flanking it denote
$P_{i\pm 1}$.
\subsection{Projective measurements}
We now discuss the so called projective or strong measurements. For this, the initial state of the apparatus is taken to be a \emph{single}
pointer state, say, $P_0$. The same measurement interaction discussed above now reads: 
\begin{equation}
|P_0\rangle_A\otimes|\pm\rangle_S\xrightarrow{{\cal M}}\,|P_\pm\rangle_A\otimes|\pm\rangle_S
\end{equation}
in an obvious relabelling of states. Henceforth we shall drop the $\otimes$.
If the initial state of the system is taken to be:
\begin{equation}
\label{eq:sysini}
|\psi\rangle = \alpha|\uparrow\rangle+\beta|\downarrow\rangle\quad\quad |\alpha|^2+|\beta|^2=1
\end{equation}
and the initial state of the apparatus-system complex 
is taken to be $|\psi\rangle\,|P_0\rangle$, the \emph{post-measurement-interaction state} of the composite is given by 
\begin{equation}
|P_0\rangle\,|\psi\rangle\xrightarrow\,|\Psi\rangle_{SA}=\alpha|P_+\rangle|\uparrow\rangle+\beta|P_-\rangle|\downarrow\rangle
\end{equation}
As is well known, this is an entangled state and does not correspond to the expected state after a definite measurement 
outcome.
The current folklore is that \emph{environmental decoherence} reduces the density matrix $\rho_{SA}$ of this
pure state to the mixed state, which by construction, is diagonal in the pointer-states bases:
\begin{equation}
\rho_{SA}\xrightarrow{decoh}\,|\alpha|^2|P_+\rangle\langle P_+|\,|\uparrow\rangle\langle \uparrow|+|\beta|^2\,|P_-\rangle\langle P_-||\downarrow\rangle\langle \downarrow|
\end{equation}
The system itself can be efficiently characterized by its \emph{reduced density matrix}:
\begin{equation}
\rho_{red} = |\alpha|^2|\uparrow\rangle\langle \uparrow|+|\beta|^2|\downarrow\rangle\langle \downarrow|\quad\quad 
\end{equation}
The so called \emph{Purity} of this mixed state, defined as $tr_A\,\rho_{SA}^2$, is given by  $ 1-2|\alpha|^2|\beta|^2$. 
This is generically far from a purity value of unity. It should be appreciated that decoherence, however, does not explain the measurement
process on an event by event basis.

With each outcome, the system is irretrievably altered. The pointer position $+1$ occurs with probability $|\alpha|^2$,
while the outcome $-1$ occurs with probability $|\beta|^2$. The mean pointer position is $|\alpha|^2-|\beta|^2$. The 
variance is the standard uncertainty associated with the state $|\psi\rangle$, and the error in the result of M
measurements falls off as $\frac{1}{{\sqrt M}}$.
\subsection{Weak measurements}
\begin{figure}[htp!]
  \centering
  \includegraphics[width=1.5in]{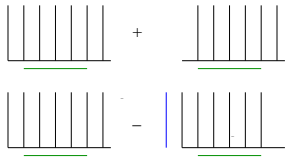}
\caption{A weak measurement}
  \label{fig:measint}
\end{figure}

Now we turn to the so called \emph{weak measurements}. To demystify the hopelessly large hype(and many wrong statements), we
consider a highly idealised example which nevertheless contains the essential features of this very interesting new category
of measurements introduced by Aharonov and his collaborators \cite{aharonovorig} (for a detailed exposition of many aspects
of weak measurements see \cite{nori}). The initial state of the apparatus is now taken to be a very
broad superposition of pointer states with equal weights and no relative phases:
\begin{equation}
|A\rangle = \frac{1}{\sqrt{N}}\,\sum_{i=1}^{i=N}\,|P_i\rangle
\end{equation}
In some of the current literature, even this very very broad state is treated as a pointer state, with its 
centroid identified as the corresponding \emph{pointer position}. It is quite meaningless to take this position.
Introducing the apparatus state
\begin{equation}
|{\bar A}\rangle = \frac{1}{\sqrt{N-2}}\,\sum_{i=2}^{i=N-1}\,|P_i\rangle
\end{equation}
one sees that the measurement interaction of eqn(\ref{eq:measint}) leads in this case to
\begin{equation}
|A\rangle|\uparrow\rangle\,\rightarrow\: \{\sqrt{\frac{N-2}{N}}\,|{\bar A}\rangle+\frac{1}{\sqrt{N}}\,
(|P_N\rangle+|P_{N+1}\rangle)\}|\uparrow\rangle
\end{equation}
\begin{equation}
|A\rangle|\downarrow\rangle\,\rightarrow\: \{\sqrt{\frac{N-2}{N}}\,|{\bar A}\rangle+\frac{1}{\sqrt{N}}\,
(|P_0\rangle+|P_{1}\rangle)\}|\downarrow\rangle
\end{equation}
This is depicted in Figure.(\ref{fig:measint}).
If the initial state of the apparatus and system is taken to be $|A\rangle\otimes|\psi\rangle$, with $|\psi\rangle$ as given by eqn.(\ref{eq:sysini}), 
the \emph{post-measurement-interaction} composite state is now given by:
\begin{eqnarray}
\label{eq:postmeasweakexample}
& &\sqrt{\frac{N-2}{N}}\,|{\bar A}\rangle|\psi\rangle+\frac{\alpha}{\sqrt{N}}\,(|P_N\rangle+|P_{N+1}\rangle)|\uparrow\rangle\nonumber\\
& &+
\frac{\beta}{\sqrt{N}}\,(|P_0\rangle+|P_1\rangle)|\downarrow\rangle
\end{eqnarray}
The \emph{post-decoherence} system-apparatus mixed state, which is by construction diagonal in $P_i$ (the incorrectness of
treating the initial apparatus state $|A\rangle$ becomes evident here), is easily worked out to be:
\begin{eqnarray}
\label{eq:postdecohweakexample}
& &\frac{N-2}{N}\,\sum_{i=2}^{i=N-1}|P_i\rangle\langle P_i||\psi\rangle\langle \psi|\nonumber\\
&+&\frac{|\alpha|^2}{N}(|P_N\rangle\langle P_N|+|P_{N+1}\rangle\langle P_{N+1}|)|\uparrow\rangle\langle \uparrow|\nonumber\\
&+&\frac{|\beta|^2}{N}(|P_0\rangle\langle P_0|+|P_{1}\rangle\langle P_{1}|)|\downarrow\rangle\langle \downarrow|
\end{eqnarray}
 The post-measurement reduced density matrix of the system is obtained by tracing over the apparatus state-space:
\begin{equation}
\label{eq:redrhoweakexample}
\rho^{weak}_{red} = |\psi\rangle\langle \psi|-\frac{2}{N}(\alpha\beta^*|\uparrow\rangle\langle \downarrow|+\alpha^*\beta|\downarrow\rangle
\langle \uparrow|)
\end{equation}
The purity of this reduced density matrix is
\begin{equation}
\label{eq:weakpurity}
{\cal P}_{weak} = 1-\frac{8}{N}|\alpha|^2|\beta|^2
\end{equation}
When $N >> 1$, this post-measurement purity can be arbitrarily close to the unit purity of the system state before
measurement. In this sense, the weak measurements appear to be highly \emph{non-invasive}, but there is more to 
invasiveness than just this measure.

A number of important properties attributed to weak measurements in general can be gleaned from this highly idealized
example. From eqn.(\ref{eq:postmeasweakexample}), it follows that with probability $1-2/N$, the system is not changed at
all(extreme weakness). It is also important to observe that this 'weakness' has nothing to do with the strength of the
measurement interaction. Rather, it is completely controlled by N, the \emph{width} of the initial apparatus state.
While with most measurement outcomes, there is no change of the system, the \emph{information} obtained about the system 
by these outcomes is also zero. This follows from the fact that the probabilities for these outcomes has \emph{no} 
dependence on the initial state. On the other hand, the outcomes $i=N, N+1$ occur with the very low probability
$\frac{|\alpha|^2}{N}$ and likewise, $i=0,1$ with probability $\frac{|\beta|^2}{N}$. For these outcomes, the system is
irretrievably changed exactly as in projective measurements! These probabilities being dependent on the system state, these
outcomes give full information! 

Let us now calculate the mean pointer position ${\bar i}$ and the associated variance. Elementary calculations give this
to be $(N+1)/2$ before measurement, and, $(N+1)/2+|\alpha|^2-|\beta|^2$. Therefore the shift in the mean pointer position 
is exactly the expectation value of the observable, as in the projective measurements. The variance in the pointer 
positions is now dramatically different. Before measurements it is $(N^2-1)/12$ while after measurements, it is still
essentially this, but shifted by a tiny system-dependent part:
\begin{equation}
\label{eq:weakexamplevariance}
(\Delta i)^2_{pre} = \frac{N^2-1}{12}\quad\quad (\Delta i)^2_{post} = (\Delta i)^2_{pre}+(\Delta S)^2_{\psi}
\end{equation}
The results for the mean and variance are exactly the same as for the most generic weak measurements \cite{nori}.  
In this elementary example, the deviations of pointer outcomes can trivially be much larger than the eigenvalues
of the observable in question. There is no big mystery that needs some special understanding. Another noteworthy 
feature is that since for pointer outcomes, the system is mostly not
an eigenstate of the observable (in the example, this happens only when the outcomes are $i=0,1,N,N+1$), there is no
\emph{value} of the observable associated with the value of the pointer outcome, unlike the case in projective measurements.
Too much has been made of this starting from the title of the first paper on weak measurements \cite{aharonovorig}.

Now we turn out to a standard treatment of weak measurements. The
 \emph{Pointer variable} is taken to be $p$ the momentum. The \emph{Pointer states} are taken to be the momentum eigenstates $|p\rangle$.
 In practice, these are taken to be narrow gaussian wave packets in momentum representation.
 As seen in our extreme example, the initial apparatus state for weak measurements should be a \emph{very broad superposition} of
pointer states i.e
\begin{equation}
\label{eq:appstategen}
|A\rangle = {\bar N}_p\,\int\,dp\,e^{-\frac{p^2}{2\Delta_p^2}}\,|p\rangle \quad\quad {\bar N}_p = (\pi\Delta_p^2)^{-1/4}
\end{equation}
with $\Delta_p >> 1$.
 The \emph{measurement interaction} is taken to be $e^{-iQA}$ where $A$ is the observable that is being measured, and Q the
variable conjugate to momentum. As in the von Neumann model, this is taken to be impulsive, acting exactly at the time of 
measurement. For simplicity, we take the observable A to have the discrete, non-degenerate spectrum 
$a_i, |a_i\rangle$.
 The initial system state is taken to be:
\begin{equation}
\label{eq:syststategen}
|\psi\rangle = \sum_i\,\alpha_i\,|a_i\rangle\quad\quad \sum_i\,|\alpha_i|^2=1
\end{equation}
 The \emph{post-measurement-interaction} state of system and apparatus is then given by:
\begin{eqnarray}
|\Psi\rangle_{SA,weak} &=& {\bar N}_p\,\sum_i\,\alpha_i\,\int\,dp\,e^{-\frac{p^2}{2\Delta_p^2}}|p+a_i\rangle|a_i\rangle\nonumber\\
&=&\,\int\,dp\,N(p,\{\alpha\})|p\rangle|\psi_p\rangle
\end{eqnarray} 
 Where
\begin{eqnarray}
N(p,\{\alpha\}) &=& {\bar N}_p\,\sqrt{\sum_i\,|\alpha_i|^2\,e^{-\frac{(p-a_i)^2}{\Delta_p^2}}}\nonumber\\ 
|\psi_p\rangle  &=& \frac{{\bar N}_p}{N(p,\{\alpha\})}\,\sum_i\,\alpha_i\,e^{-\frac{(p-a_i)^2}{2\Delta_p^2}}|a_i\rangle
\end{eqnarray}
Hence, weak measurements can be viewed as the so called Positive Operator Valued Measurements(POVM) with 
measurement operators:
\begin{equation}
\label{eq:weakpovm}
M_p = {\bar N}_p\,\sum\,e^{-\frac{(p-a_i)^2}{2\Delta_p^2}}|a_i\rangle\langle a_i|
\end{equation}
 The \emph{post-decoherence} mixed state of the system and apparatus is easily calculated to be:
\begin{equation}
\rho^{post-decoh}_{SA} = \int\,dp\,|N(p,\{\alpha\})|^2\, |p\rangle\langle p||\psi_p\rangle\langle \psi_p|
\end{equation}
 The probability distribution for the pointer outcomes is given by $|N(p,\{\alpha\}|^2$. As the eigenvalues $a_i$ are
bounded, this distribution, when $p|a_i| << \Delta_p^2$, is well approximated by
\begin{equation}
\label{eq:lowpapprox}
|N(p,\{\alpha\}|^2\simeq\quad {\bar N}_p^2\,e^{-\frac{P^2}{\Delta_p^2}}+\ldots
\end{equation}
In this case
\begin{equation}
\label{eq:lowpstate}
|\psi_p\rangle\simeq\,|\psi\rangle+\ldots
\end{equation}
where the dots represent small corrections. One once again observes the same features encountered in the example, namely,
that for most of the outcomes the state changes very little(in the example, that change was zero while in 
the more
realistic cases, as here, it is small). But precisely for those cases, the probability of outcome is either 
independent, or nearly independent, of the system
state and no information can be obtained about the system state. Nevertheless, as in the example, the mean pointer position
has full information about the state(provided a \emph{complete} set of weak measurements are performed).
 The average outcome and its variance can be calculated exactly:
\begin{equation}
\langle p \rangle = \sum_i\,|\alpha_i|^2\,a_i = \langle A \rangle_\psi\quad (\Delta p)^2 = \frac{\Delta_p^2}{2}+(\Delta A)^2_\psi
\end{equation}
 These necessarily have to be \emph{ensemble} measurements.
 The errors in weak measurements are very large because $\Delta_p >> 1$. These are to be reduced \emph{statistically}. 
It is instructive to compute the \emph{reduced} density matrix of the system:
\begin{equation}
\label{eq:weakredrho}
\rho^{red,weak}_{sys} = |\psi\rangle\langle \psi| -\frac{1}{4\Delta_p^2}\sum_{i,j}\,\alpha_i\alpha_j^*\,(a_i-a_j)^2|a_i\rangle\langle a_j|
\end{equation}
\section{Weak Measurements and Leggett-Garg Inequalities}
 We saw in eqn.(\ref{eq:weakredrho}) that the reduced density matrix after a weak measurement is \emph{practically} the same as the initial pure density matrix.
 In this sense, the weak measurements can be said to be \emph{non-invasive}.
 Non-invasive measurements have been emphasized in a variety of contexts.
 The most notable of these has been the \emph{Leggett-Garg} inequalities \cite{agarg,dhome,mahesh}.
 A typical experimental setup
 consists of four series of measurements on
identical initial states.
 In each series, some quantity $Q(t)$ is measured at two instants of time. In the first, measurements are done at $t_1$ and $t_2$;
in the second, at $t_2$ and $t_3$, in the third at $t_3$ and $t_4$, and finally in the fourth at $t_1$ and $t_4$. It is to be noted
that $t_1 < t_2 < t_3 < t_4$.

 The first measurement in each series is required to be \emph{non-invasive}, as then the second measurement can be 
\emph{construed} to have also been made on the \emph{same state} as the initial one.
 Thus a total of 8 measurements of which 4 have to be non-invasive.
 The natural question is whether weak measurements can be used to achieve this?
 The answer to this hinges on the accuracy of measurements(errors) as well as the \emph{available resources}, the apparent
non-invasiveness of weak measurements notwithstanding.
 An obvious resource to be considered is the \emph{ensemble size} of the initial state. Let this be M identical copies.

 If we consider using weak measurements to provide the required non-invasive measurements, it will be necessary to divide
M into 4 equal subensembles of $M/4$ copies each, and use one for each series of measurements.
 The statistical error in the resulting weak measurements will be $\epsilon_w=\frac{\Delta_p}{\sqrt{2}}\frac{1}{\sqrt{M/4}}$.
 It should be remembered that for the second measurement in each series, the state will not be exactly the same as the
original state. Depending on $\Delta_p$, this could be an important factor to reckon with in practical implementations.
 Since the second measurement does not have to be non-invasive, it can even be done with strong measurements, which, for
the same ensemble size would yield an error substantially lowered by a factor $\frac{\sqrt{2}(\Delta A)_\psi}{\Delta_p}$.
 The error analysis of LG-inequalities would be more complicated then.

 Let us estimate the ensemble size that would yield the same error $\epsilon_w$ but now done with strong measurements.
 The relation between statistical error and ensemble size for strong measurements is $\epsilon_s = \frac{(\Delta A)_\psi}{\sqrt{M_s}}$,
where $M_s$ is the relevant ensemble size.
 Therefore the ensemble size for strong measurements with the same error as in the weak measurements is:
\begin{equation}
M_s = \frac{(\Delta A)_\psi^2}{\epsilon_w^2}=\frac{M}{2}\cdot\frac{(\Delta A)_\psi^2}{\Delta_p^2}
\end{equation}
 The idea now is to divide the original resource into 8 equal subensembles and use each of them to perform the total of 8 measurements
required.

 Altogether 8 strong measurements need to be done and the total ensemble size required is $M\cdot\frac{4(\Delta A)_\psi^2}{\Delta_p^2}$
 Hence it follows that as long as $\frac{(\Delta A)_\psi}{\Delta_p} << 1/2$, the ensemble size required for strong version of checking LG inequalities is \emph{much smaller} than what was required for the weak version of the same!
 Furthermore, in the strong version the states used for all the 8 measurements are \emph{exactly identical}!
 In summary, if $\Delta_p$ is very large, one can test the LG-inequalities with much smaller resources using strong measurements.
 If $\Delta_p$ is not so large, the weak measurements are no longer non-invasive.
 Either way, there is no case for invoking weak measurements to test the LG-inequalities. Similar considerations for
determination of so called trajectories will be taken up elsewhere.

\section{Repeated Weak Measurements On a Single Copy}
 One of the most \emph{surprising} and \emph{shocking} facets of the Copenhagen view of quantum mechanics is what 
one may call the \emph{demise of the individual} (for a detailed exposition see \cite{ndhonto}).
 More precisely, that view predicated that no information can be obtained about the unknown state of a \emph{single copy}.
 This is a trivial consequence if one uses projective or strong measurements.
 This is so as the first measurement randomly results in an eigenstate and all subsequent measurements have no bearing on the
original unknown state.
 Weak measurements offer a \emph{superficial hope} that it may be possible to determine the unknown state of a single copy.
 The basis for that hope is that each weak measurement, with high probability, very weakly alters the system state while giving
some information about the original state.

Consider the following \emph{schema} for repeated weak measurements on a single copy. (i) Perform a weak measurement 
of observable A on a single copy of an unknown state $|\psi\rangle$.  
 Let the apparatus outcome be, say, $p_1$.
 Consequently, the system state at this stage is $|\psi_{p_1}\rangle$.
 (ii) Restore the apparatus to the same state before the first weak measurement.
 (iii) Perform weak measurement of A in the new system state $|\psi_{p_1}\rangle$.
 (iv) Repeat.

The crucial question is whether the statistics of outcomes $p_1,p_2,...,p_N$ have anything to say about the 
original unknown $|\psi\rangle$? The naive argument would be that since at each step one gathers some \emph{information}
about the original unknown state, although very little, with sufficiently large repetitions one ought to gather enough
information to determine the original state. The question will be answered in the negative here. The details can be found in \cite{ndhweakrepeat}. Alter and Yamamoto \cite{orly,orlybook} had
in fact analysed a very similar problem in the context of repeated QND measurements long ago, but issues of degradation of the state 
as well connections 
to strong measurements were not considered by them. 

The probability $P^{(1)}(p_1)$ of the first outcome $p_1$ is given by:
\begin{equation}
P^{(1)}(p_1) = |N(p_1,\{\alpha\}|^2
\end{equation}
 The system state after this outcome is $|\psi_{p_1}\rangle$.
 It is useful to describe this state as one with \emph{changed values} of $\{\alpha\}$:
\begin{equation}
\alpha_i^{(1)} = \frac{{\bar N}_p}{N(p_1,\{\alpha\})}\:e^{-\frac{(p_1-a_i)^2}{2\Delta_p^2}}
\end{equation}
 The probability $P(p_2)$ of the second outcome $p_2$ is, therefore:
\begin{eqnarray}
& &P^{(2)}(p_2) = |N(p_2,\{\alpha^{(1)}\}|^2\nonumber\\
&=&\frac{{\bar N}_p^4}{N(p_1,\{\alpha\})^2}\,
\sum_i\,|\alpha_i|^2\,e^{-\frac{(p_1-a_i)^2}{\Delta_p^2}}\cdot e^{-\frac{(p_2-a_i)^2}{\Delta_p^2}}
\end{eqnarray}

 But this is the \emph{conditional probability} $P(p_2|p_1)$ for obtaining $p_2$ \emph{given} that the first outcome was $p_1$.
 The \emph{joint probability} distribution $P(p_1,p_2)$ is given by Bayes theorem to be $P(p_1)\cdot P(p_2|p_1)$:
\begin{equation}
P(p_1,p_2) = {\bar N}_p^4\,
\sum_i\,|\alpha_i|^2\,e^{-\frac{(p_1-a_i)^2}{\Delta_p^2}}\cdot e^{-\frac{(p_2-a_i)^2}{\Delta_p^2}}
\end{equation}
 The state of the system when the outcomes are $p_1,p_2$ is:
\begin{equation}
|\psi(p_1,p_2)\rangle = \frac{\sum_i\,\prod_{j=1}^2\,e^{-\frac{(p_j-a_i)^2}{\Delta_p^2}}\alpha_i\,|a_i\rangle}{\sqrt{\sum_i\,
\prod_j\,|\alpha_i|^2\,e^{-\frac{(p_j-a_i)^2}{\Delta_p^2}}}}
\end{equation}

 These readily generalize to the case of M repeated measurements:
\begin{eqnarray}
P(p_1,p_2,\ldots,p_M) &=& ({\bar N}_P^2)^M\,\sum_i\,|\alpha_i|^2\,\prod_{j=1}^M\,e^{-\frac{(p_j-a_i)^2}{\Delta_p^2}}\nonumber\\
|\psi(p_1,p_2,..,p_M)\rangle &=& \frac{\sum_i\,\prod_{j=1}^M\,e^{-\frac{(p_j-a_i)^2}{\Delta_p^2}}\alpha_i\,|a_i\rangle}{\sqrt{\sum_i\,
\prod_j\,|\alpha_i|^2\,e^{-\frac{(p_j-a_i)^2}{\Delta_p^2}}}}
\end{eqnarray}
 These equations codify \emph{all} the information that can be obtained by repeated weak measurements on a single copy of an unknown state.
 The joint probability distribution is not \emph{factorisable} as the outcomes are \emph{not mutually independent}, but it is still of
the so called \emph{separable} form.

 The average $y_M$ of the M outcomes is $\sum_i\,|\alpha_i|^2\,a_i = \langle A \rangle_\psi$!
 Does this mean we have obtained the same information in a weak measurement on a single copy what
could only be obtained by ensemble measurements of the strong kind?
 It is necessary to look into the distribution function $P(y_M)$ for such an average.
 Recall that in ensemble measurements this takes the form (Central Limit Theorem):
\begin{equation}
P(y_M)_{ensemble} = {\tilde N}\,e^{-\frac{M(y_M-\mu)^2}{\Delta^2}}
\end{equation}
 In ensemble measurements too, the sequence of outcomes in \emph{a particular realization} will be different, and \emph{unpredictable}.
 But the average obtained in any particular realisation converges to the true average as 
$M\rightarrow\,\infty$. 
 Now it turns out that the story is entirely different for repeated weak measurements on a single copy!

 The distribution function $P(y_M)$:
\begin{eqnarray}
P(y_M)&=& \sqrt{\frac{M}{\pi\Delta_p^2}}\,\sum_i\,|\alpha_i|^2\,e^{-\frac{M(y_M-a_i)^2}{\Delta_p^2}}\nn\\
&\rightarrow&
\sum_i\,|\alpha_i|^2\,\delta(y_M-a_i)
\end{eqnarray}
 The distribution of $y_M$ is no longer peaked at the true average with errors decreasing as $M^{-1/2}$.
 Instead, it is a weighted sum of distributions that increasingly peak around the eigenvalues as $\Delta_p$ increases.
 In the limiting case, averages over a particular realisation will be eigenvalues occurring with probability $|\alpha_i|^2$, exactly
as in the case of strong measurements.
 Hence averages over any particular realisation do not give any information about the initial state.

 To substantiate this picture further, one can investigate the average value of the post-measurement system reduced density
matrix:
\begin{equation}
\langle \rho^{red} \rangle = \rho - \sum_{i,j}\,\alpha_i\alpha_j^*\,(1-e^{-\frac{M(a_i-a_j)^2}{4\Delta_p^2}})|a_i\rangle\langle a_j|
\end{equation}
 Therefore as M becomes larger and larger, there is significant change in the system state.
 In the limit $M\rightarrow\,\infty$, the \emph{off-diagonal} parts of the density matrix get completely quenched, as
in \emph{decoherence}!

 In that limit, the density matrix becomes diagonal in the eigenstate(of A) basis:
\begin{equation}
\langle \rho^{red} \rangle\,\rightarrow\: \sum_i\,|\alpha_i|^2\,|a_i\rangle\langle a_i|
\end{equation}
 This is exactly the post-measurement system state in the case of strong measurements.
 It should be noted that this decoherence in eigenstate basis has nothing to do with the environmental decoherence
in the pointer state basis of the apparatus.
 It is entirely due to the large number of repeated weak measurements.
 Such an effect had also been noted by Gurvitz in 1997 \cite{gurwitz}.

 We can view the distance between the initial $\rho$ and the average post-measurement reduced density matrix
$\langle \rho^{red} \rangle$, according to some reasonable distance measure, as a measure of the \emph{disturbance}
caused by the repeated weak measurements on the single copy.
 For example, ${\cal D} = 1-tr\,\rho\,\langle \rho^{red}\rangle$ is one such distance measure.
 The statistical error $\epsilon = \frac{\Delta_p}{\sqrt{2M}}$.
 Then one gets the \emph{error-disturbance} relation:
\begin{eqnarray}
{\cal D}(\epsilon)&=& \sum_{i,j}\:|\alpha_i|^2|\alpha_j|^2(1-e^{-\frac{(a_i-a_j)^2}{8\epsilon^2}})\nn\\
&\rightarrow&
\sum_i\,|\alpha_i|^2(1-|\alpha_i|^2)
\end{eqnarray}
 Reducing errors can only be at the cost of increasing invasiveness! It should be noted that this error-disturbance relation bears
no obvious relation to the ones being discussed by Ozawa \cite{ozawa}.
\section{Weak value coordinates and optimal weak value measurements}
This section is based on the works \cite{ndhsaicoord} and \cite{ndhsaioptimal}.
 Consider the projection operators ${\cal P}_\pm$ for the eigenstates $|\pm\rangle$ of, say, $S_z$. 
 Let the preselected state be $|\psi =\alpha |+\rangle\,+\,\beta|-\rangle$, with $|\alpha|^2+|\beta|^2=1$.
 Let the post-selected state be $|b\rangle$.
 If $w_\pm$ are the weak values of ${\cal P}_\pm$
\begin{equation}
w_\pm=\frac{\langle b||\pm\rangle\langle \pm||\psi\rangle}{\langle b|\psi\rangle} \quad\quad w_+\,+\,w_-=1
\end{equation}
 The idea of \emph{weak value tomography}($b_\pm=\langle b|\pm\rangle$):
\begin{equation}
\alpha = \frac{\frac{w_+}{b_+}}{\sqrt{|\frac{w_+}{b_+}|^2+|\frac{w_-}{b_-}|^2}}\quad\quad 
\beta = \frac{\frac{w_-}{b_-}}{\sqrt{(\frac{w_+}{b_+})^2+(\frac{w_-}{b_-})^2}}
\end{equation}
 Thus experimentally determining a single complex weak value ($w_+$ or $w_-$) suffices to determine the state.
 $w_+=\frac{1}{2}+w_z$ and $w_-=\frac{1}{2}-w_z$, where $w_z$ is the weak value of $S_z$.
 Thus it suffices to measure the weak value of a single observable to determine the state as against conventional tomography
which would require the \emph{expectation values} of \emph{two} independent observables and a \emph{sign}!
 At this stage, the fact that $Re\: w, Im\: w$ are \emph{unbounded} becomes crucial.
 It indicates that the real and imaginary parts of weak values provide a \emph{stereographic projection} of the \emph{Riemann sphere}.

 The \emph{metric} on the state space can be introduced through the line element
\begin{equation}
dl^2 = 2\,tr\,d\rho\,d\rho
\end{equation}
 For example, if the pure state density matrix is parametrised as
\begin{equation}
\rho = \frac{I}{2} +\langle S_x \rangle\,\sigma_x +\langle S_y \rangle\,\sigma_y +\langle S_z \rangle\,\sigma_z
\end{equation}
with
\begin{equation}
\langle S_x \rangle^2+ \langle S_y \rangle^2+ \langle S_z \rangle^2 = \frac{1}{4}
\end{equation}
 The line element becomes
\begin{equation}
dl^2 = 4\{(dS_x)^2+(dS_y)^2+(dS_z)^2\}
\end{equation}
 This is just the metric on a sphere.

 The most general form of the line element is
\begin{equation}
dl^2 = g_{ww}\,dw^2\,+g_{{\bar w}{\bar w}}\,d{\bar w}^2\,+g_{w{\bar w}}\,dw\,d{\bar w}
\end{equation} 
 Explicit evaluation yields
\begin{equation}
g_{w{\bar w}}=\frac{4}{|b_+|^2|b_-|^2}\,\frac{1}{\sqrt{|\frac{w_+}{b_+}|^2+\frac{w_-}{b_-}|^2}}
\end{equation}
with $g_{ww} = g_{{\bar w}{\bar w}}=0$.
 Therefore, the  weak value coordinates have the nice feature that they are \emph{conformal}!

 In terms of $Re\:w_+=x,Im\:w_+=y$, the line element can be rewritten as
\begin{equation}
dl^2 = \frac{4|b_+|^2|b_-|^2\:(dx^2+dy^2)}{\{x^2+y^2+x(|b_-|^2-|b_+|^2)+\frac{1}{4}\}^2}
\end{equation}
 The volume(area) element of the state space is then
\begin{equation}
dA = \frac{4|b_+|^2|b_-|^2\:dx\,dy}{\{x^2+y^2+x(|b_-|^2-|b_+|^2)+\frac{1}{4}\}^2}
\end{equation}
 The total volume of $\rho$-space is correctly reproduced as \emph{4$\pi$}(area of unit sphere).

 Now another remarkable feature of weak measurements comes into play i.e the measurement errors in both x and y are the
\emph{same}, and are \emph{state-independent}.
 The common statistical error is $\Delta_s = \frac{\Delta_p}{\sqrt{2M}}$.
 This is in contrast to strong measurements.
 Following Wootters and Fields, the \emph{error volume} is
\begin{equation}
(\Delta A)_{err} = \frac{16\,\Delta_s^2\,|b_+|^2|b_-|^2\:dx\,dy}{\{x^2+y^2+x(|b_-|^2-|b_+|^2)+\frac{1}{4}\}^2}
\end{equation}
 As noted by Wootters and Fields in the case of standard tomography, this is \emph{state-dependent}, and it is not possible to
optimise it.
 We follow them and optimise the error volume \emph{averaged over state space}.

 The state averaged error volume can easily be worked out:
\begin{equation}
\langle (\Delta A)_{err} \rangle = \frac{16\,\Delta_s^2}{|b_+|^2|b_-|^2}
\end{equation}
 Since $\Delta_s$ has no dependence on the post-selected state $|b\rangle$, it is straight forward to optimise this.
 The solution is $|b_+|^2=|b_-|^2=\frac{1}{2}$.
 In other words \emph{weak value measurements are optimal in the sense of minimizing state averaged error volume when the
post-selected states are MUB with respect to the eigenstates of the observable measured}.
 Extension to spin-1 and higher spin values is under investigation.

\vspace{2mm}
\noindent{ACKNOWLEDGEMENTS: I thank Dipankar Home and T.S. Mahesh for discussions on non-invasive measurements. I also acknowledge support from DST, India, for the project IR/S2/PU-801/2008.}

\end{document}